\documentstyle[floats,twocolumn,aps,prl] {revtex}
\begin{document}
\draft
\tighten

\noindent {\bf Gross and Scheffler Reply:} 
\noindent Among other things (e.g. steering and steric effects in
dissociative
adsorption) we had predicted~\cite{Gro95}
that the initial sticking probability of H$_2$ molecules
impinging at clean Pd\,(100) exhibits oscillations,
reflecting the quantum nature of the scattering process. In the preceding
comment Rettner and Auerbach (RA) analyze experimental results and
conclude that these oscillations are not detectable and thus 
either not existing or at least very small.

In this reply we  argue that the experimental study of RA
is not conclusive to rule out the existence of quantum oscillations in the
scattering of H$_2$ and note several problems and 
incongruities in their study:\\[-0.7cm]
\begin{itemize}
   \item[1)] In their figure RA compare measurements at an angle of 
incidence $\theta_i = 15^{\circ}$ to our 
calculations performed for normal-incidence, i.e.
$\theta_i = 0^{\circ}$~\cite{Gro95}.\\[-0.7cm]
   \item[2)] RA argue  in their analysis that substrate
vibrations can be treated in a ``surface mass model''.\\[-0.7cm]
   \item[3)]  The experimental intensity of reflected H$_2$
molecules was integrated over a large angle.\\[-0.7cm]
\end{itemize}
We now elaborate on these points and explain  
some aspects which show that the measurement of
the predicted quantum oscillations is a most challenging project.

{\em 1) Angle of incidence} -- The oscillatory structure of the sticking
probability is a consequence of the quantum nature of H$_2$ scattering
and reflects the opening of new scattering channels and resonances with
increasing kinetic energy. In a simplified
description, i.e. neglecting rotational and vibrational degrees
of freedom of H$_2$, elastic scattering
gives rise to reflected beams: $|({\bf k}_{\parallel} + {\bf g}),
- \sqrt{ k_{z}^{2} - 2{\bf k}_{\parallel} \cdot {\bf g}
- {\bf g}^{2} } \ \rangle$.
Here $({\bf k}_{\parallel}, k_{z})$ is the wave vector of the 
incident H$_2$ beam and {\bf g} is a two-dimensional
reciprocal lattice vector of the surface. The condition
for emerging beams is that the argument under the square root
is positive. Just before a new beam can emerge, 
it is already built up though it remains confined to the surface.
Thus, this beam can not (yet) be observed directly, but as it is
coherent with the other beams these other
will exhibit sharp resonance structures. Furthermore, oscillations
could be caused by selective adsorption resonances \cite{stern}.
If there are  many scattering states, the effect will be small, 
because it will be distributed over all states. 

Hence the strengths and energies of the oscillatory
structures depend sensitively on the initial conditions.
A general angle of incidence has disadvantages because the number of
symmetrically distinct states is larger than for normal incidence.
For an incident angle of $\theta_i = 15^{\circ}$ the oscillations
are therefore much shallower compared to the normal-incidence results,
as confirmed by recent calculations that we have performed.
-- These effects are well known  from other quantum-mechanical
scattering studies, e.g. in low energy electron diffraction (see, e.g.
Ref.~\cite{mcrae,pendry}).

{\em 2) Influence of substrate vibrations} --  As discussed under item 1,
the oscillatory effects are due to molecules which occupy beams
which are still trapped at the surface, i.e. which are
bouncing back and forth between the substrate and the energy barrier
towards the vacuum. This dynamical trapping will lead to a much larger 
influence of the substrate vibrations than considered in the 
``surface mass model'', which simply averages over incident velocities.
In fact, not the phonon energies are expected to cause 
the main problem, but the loss of coherence of the 
temporarily trapped hydrogen.

{\em 3) Integrating over all directions} --  Such integration typically
obstructs the observation of quantum oscillations because it also 
considers all incoherently scattered molecules.
In fact, for a slightly imperfect Pd\,(100) surface it is likely that most
molecules are reflected into ``inelastic directions'' $({\bf k}_{\parallel} +
{\bf g} + {\bf q})$, where
{\bf q} is a wave vector describing the interaction with a phonon or
with a surface imperfection. 
The sticking probability of H$_2$ at small kinetic energies of $\leq 0.05$~eV
is rather large ($\approx 60 \%$).
This leads very rapidly to the adsorption of some hydrogen during scattering 
experiments, a particularly strong effect at low surface temperatures as
used by RA.  From the above discussion it can be inferred that the 
oscillatory structure is particularly sensitive to the surface potential,
and even a small number of adatoms and other surface imperfections (e.g.
steps) will reduce the scattering coherence.
As coherence is lost, these
molecules will not contribute to the oscillatory behavior.
--  These incoherence effects will be largely filtered out if only
well defined quantum states, which are consistent with elastic scattering, 
e.g. one reflected beam, were measured.
We therefore suggest to monitor only
{\em diffraction} intensities and not the whole reflection flux
in order to resolve an oscillatory structure.

In conclusion, we are convinced that the quantum mechanical resonance 
structures of H$_2$ dissociation and scattering do exist,
but the experimental detection certainly requires particular care.
Under appropriate experimental conditions the predicted oscillations
will be observable, as they have
been found in He and H$_2$ scattering since the 1930s \cite{stern} and 
in electron scattering~\cite{mcrae,pendry}.

We thank  K. Kambe, K.D. Rendulic and C. Stampfl
for helpful comments.

Axel Gross and Matthias Scheffler\\
Fritz-Haber-Institut der Max-Planck-Gesellschaft\\
Faradayweg 4-6\\
D-14195 Berlin-Dahlem, Germany\

\vspace{-0.4cm}
\begin{thebibliography}{99}
\vspace{-1.2cm}
\bibitem{Gro95} A. Gross, S. Wilke, and M. Scheffler,
Phys. Rev. Lett. {\bf 75}, 2718 (1995).
\bibitem{stern} R. Frisch and O. Stern, Z. Phys. {\bf 84}, 430 (1933).
\bibitem{mcrae} E. G. McRae, Rev. Mod. Phys. {\bf 51}, 541 (1979).
\bibitem{pendry}  J. B. Pendry, {\it Low energy electron diffraction}, 
Academic Press, London (1974), p. 112.
\end{thebibliography}
\end{document}